\documentclass[twoside,twocolumn,9pt]{article}
\usepackage{extsizes}
\usepackage[super,sort&compress,comma]{natbib}

\usepackage[version=3]{mhchem}
\usepackage[left=1.5cm, right=1.5cm, top=1.785cm, bottom=2.0cm]{geometry}
\usepackage{balance}
\usepackage{mathptmx}
\usepackage{sectsty}
\usepackage{graphicx}
\usepackage{lastpage} 

\usepackage[format=plain,justification=justified,singlelinecheck=false,font={stretch=1.125,small,sf},labelfont=bf,labelsep=space]{caption}
\usepackage{float}
\usepackage{fancyhdr}
\usepackage{fnpos}
\usepackage[english]{babel}
\addto{\captionsenglish}{%
  
}
\usepackage{array}
\usepackage{droidsans}
\usepackage{charter}
\usepackage[T1]{fontenc}
\usepackage[usenames,dvipsnames]{xcolor}
\usepackage{setspace}
\usepackage[compact]{titlesec}
\usepackage[hidelinks]{hyperref}
\usepackage{makecell}
\usepackage{color}

\usepackage{xcolor}
\usepackage{soul}

\newcommand{\V}{\textbf{V}}
\newcommand{\J}{\textbf{J}}
\newcommand{\E}{\textbf{E}}
\newcommand{\B}{\textbf{B}}
\newcommand{\D}{\textbf{D}}
\newcommand{\I}{\mathrm{I}}
\newcommand{\Det}{\mathrm{Det}}

\newcommand{\bigO}{\mathcal{O}}
\newcommand{\brho}{\mbox{\boldmath{$\rho$}}}
\newcommand{\bnabla}{\mbox{\boldmath{$\nabla$}}}
\newcommand{\bea}{\begin{eqnarray}}
\newcommand{\eea}{\end{eqnarray}}
\usepackage{hyperref}
\hypersetup{
     colorlinks = true,
     citecolor  = blue,  
     linkcolor  = magenta 
}

\usepackage{soul}

\usepackage{amsmath}
\usepackage{amssymb}
\DeclareRobustCommand\hbar{{\mathchar'26\mkern-9muh}}

\usepackage{epstopdf}

\usepackage{multirow} 

\global\arraycolsep=2pt
\usepackage{stmaryrd}
\usepackage{amsmath}
\usepackage{amssymb}
\usepackage{graphicx}
\usepackage{textcomp}
\usepackage{calrsfs}
\usepackage{caption}
\usepackage{subcaption}
\usepackage{yfonts}
\usepackage{bm}
\usepackage{soul}
\usepackage{color}
\newcommand{\ben}{\begin{equation*}}
\newcommand{\een}{\end{equation*}}
\newcommand{\bean}{\begin{eqnarray*}}
\newcommand{\eean}{\end{eqnarray*}}

\newcommand{\be}{\begin{equation}}
\newcommand{\ee}{\end{equation}}

\definecolor{cream}{RGB}{222,217,201}

\usepackage{soul}

\begin{document}

\pagestyle{fancy}
\thispagestyle{plain}
\fancypagestyle{plain}{
\renewcommand{\headrulewidth}{0pt}
}

\makeFNbottom
\makeatletter
\renewcommand\LARGE{\@setfontsize\LARGE{15pt}{17}}
\renewcommand\Large{\@setfontsize\Large{12pt}{14}}
\renewcommand\large{\@setfontsize\large{10pt}{12}}
\renewcommand\footnotesize{\@setfontsize\footnotesize{7pt}{10}}
\makeatother

\renewcommand{\thefootnote}{\fnsymbol{footnote}}
\renewcommand\footnoterule{\vspace*{1pt}%
\color{cream}\hrule width 3.5in height 0.4pt \color{black}\vspace*{5pt}}
\setcounter{secnumdepth}{5}

\makeatletter
\renewcommand\@biblabel[1]{#1}
\renewcommand\@makefntext[1]%
{\noindent\makebox[0pt][r]{\@thefnmark\,}#1}
\makeatother
\renewcommand{\figurename}{\small{Fig.}~}
\sectionfont{\sffamily\Large}
\subsectionfont{\normalsize}
\subsubsectionfont{\bf}
\setstretch{1.125} 
\setlength{\skip\footins}{0.8cm}
\setlength{\footnotesep}{0.25cm}
\setlength{\jot}{10pt}
\titlespacing*{\section}{0pt}{4pt}{4pt}
\titlespacing*{\subsection}{0pt}{15pt}{1pt}

\fancyfoot{}
\fancyfoot[LO,RE]{\vspace{-7.1pt}\includegraphics[height=9pt]{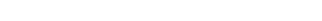}}
\fancyfoot[CO]{\vspace{-7.1pt}\hspace{11.9cm}\includegraphics{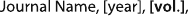}}
\fancyfoot[CE]{\vspace{-7.2pt}\hspace{-13.2cm}\includegraphics{head_foot/RF}}
\fancyfoot[RO]{\footnotesize{\sffamily{1--\pageref{LastPage} ~\textbar  \hspace{2pt}\thepage}}}
\fancyfoot[LE]{\footnotesize{\sffamily{\thepage~\textbar\hspace{4.65cm} 1--\pageref{LastPage}}}}
\fancyhead{}
\renewcommand{\headrulewidth}{0pt}
\renewcommand{\footrulewidth}{0pt}
\setlength{\arrayrulewidth}{1pt}
\setlength{\columnsep}{6.5mm}
\setlength\bibsep{1pt}

\makeatletter
\newlength{\figrulesep}
\setlength{\figrulesep}{0.5\textfloatsep}

\newcommand{\topfigrule}{\vspace*{-1pt}%
\noindent{\color{cream}\rule[-\figrulesep]{\columnwidth}{1.5pt}} }

\newcommand{\botfigrule}{\vspace*{-2pt}%
\noindent{\color{cream}\rule[\figrulesep]{\columnwidth}{1.5pt}} }

\newcommand{\dblfigrule}{\vspace*{-1pt}%
\noindent{\color{cream}\rule[-\figrulesep]{\textwidth}{1.5pt}} }

\makeatother

\twocolumn[
  \begin{@twocolumnfalse}
{\includegraphics[height=30pt]{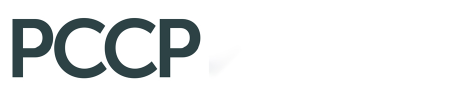}\hfill\raisebox{0pt}[0pt][0pt]{\includegraphics[height=55pt]{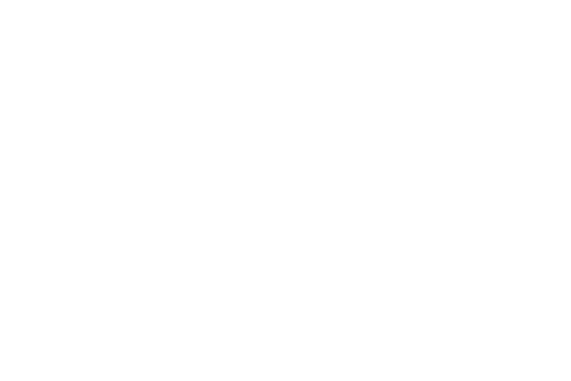}}\\[1ex]
\includegraphics[width=18.5cm]{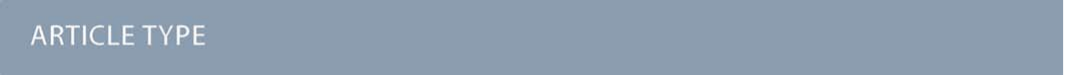}}\par
\vspace{1em}
\sffamily
\begin{tabular}{m{4.5cm} p{13.5cm} }

\includegraphics{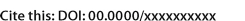} & \noindent\LARGE{\textbf{Dispersion Interaction Between Thin Conducting Cylinders}} \\
\vspace{0.3cm} & \vspace{0.3cm} \\

 & \noindent\large{Subhojit Pal,$^{\ast}$\textit{$^{a,b}$} Iver Brevik,\textit{$^{c}$} and Mathias Bostr\"om$^{\ast}$\textit{$^{a,b}$}} \\

\includegraphics{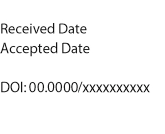} & \noindent\normalsize{The ground state and excited state resonance dipole-dipole interaction energy between two elongated conducting molecules are explored.  We review the current status for ground state interactions. This interaction is found to be of a  much longer range than in the case when the molecules are pointlike and nonconducting. 
These are well known results found earlier by Davies, Ninham, and Richmond, and later, using a different formalism, by Rubio and co-workers. We show how the theory can be extended to excited state interactions.
A characteristic property following from our calculation is that the interaction energy dependence with separation ($R$) goes like $f(R)/R^2$ both for resonance and for the van der Waals case in the long range limit. In some limits $f(R)$ has a logarithmic dependency and in others it takes constant values. We predict an unusual slow decay rate for the energy transfer between conducting molecules.} \\
\end{tabular}

 \end{@twocolumnfalse} \vspace{0.6cm}

  ]

\renewcommand*\rmdefault{bch}\normalfont\upshape
\rmfamily
\section*{}
\vspace{-1cm}


\footnotetext{\textit{$^{a}$~Centre of Excellence ENSEMBLE3 Sp. z o. o., Wolczynska Str. 133, 01-919, Warsaw, Poland;  E-mail: subhojit.pal@ensemble3.eu; mathias.bostrom@ensemble3.eu}}
\footnotetext{\textit{$^{b}$~Chemical and Biological Systems Simulation Lab, Centre of New Technologies, University of Warsaw, Banacha 2C, 02-097 Warsaw, Poland. }}
\footnotetext{\textit{$^{c}$~Department of Energy and Process Engineering, Norwegian University of Science and Technology, NO-7491 Trondheim, Norway }}

\footnotetext{\textit{$^{\ast}$~Corresponding Authors}}





\section{Introduction}

{The van der Waals force},\,\cite{Bordagbook,Ninhb,Ser2018} {exhibits a long-range nature, acts between  atoms or molecules that are potentially kept at very large separation, both when these particles have permanent moments and when they don't}.  The force is attractive, and is entirely quantum mechanical in nature since it depends on $\hbar$. The force is however nonrelativistic  since it does not contain the velocity of light $c$. The central physical element in this effect is the occurrence of fluctuating electromagnetic fields,  which induce instantaneous dipole moments in the particles.  The theory of the van der Waals force was developed by London in 1930\,\cite{london1930theorie}.
 A related phenomenon is the resonance interaction between a pair of atoms or molecules in an excited energy state.
The resonant energy transfer was discovered experimentally by Cario and Franck in 1923\,\cite{Cario} and explored theoretically by F\"{o}rster\,\cite{Forster}. It is important in biology and biophysics, e.g. for photosynthesis, light harvesting, and fluorescent light-emitting devices\,\cite{Groendelle1}. It can also be used to create entangled states for quantum logic using both molecules\,\cite{Brennen} and quantum dots\,\cite{Perina}. Resonance interaction has also been exploited to create cold molecules\,\cite{Jones}.   The underlying QED perturbation theory\,\cite{Stephen,McLachlan,Mclone} and a self-consistent semi-classical theory\,\cite{Bostrom1} both provide the same general expressions for the non-retarded resonance interaction energy between non-conducting molecules.  In order to derive correct results, in the retarded finite temperature limit, a self-consistent formalism is 
 essential\,\cite{Bostrom1}.
 {It has in the past been assumed that atom-based theories give simple power laws dependence on separation (e.g. for spheres, cylinders and thin plates) which only depends on the independent directions the objects extend. However, recent atom-based work by Dobson}\,\cite{dobson2023mbd+}  {has predicted $R^{-2}(\ln(R/a))^{-3/2}$ dispersion energy for elongated conducting molecules.} For the case with thin non-conducting cylinders the non-retarded interaction decays as $1/R^5$ for large distances\,\cite{mitchell1973van,langbein1972van,ambrosetti2016wavelike} and {similar power law dependency was found by Misquitta }{\it et al.}\,\cite{misquitta2010dispersion} { for semiconducting nanowires with zero band gap at large inter wire separations}.

We find substantial new physics when considering a pair of elongated (cylindrical) conducting molecules in the non-retarded limit. 
The ground state van der Waals interaction between a pair of thin elongated conducting molecules is relevant for gold nanoparticles in nanotechnology\,\cite{OLIVEIRA2021768} and for DNA-DNA interaction in biotechnology. {Angyan et al., in their book ``London Dispersion Forces”, discussed long-ranged dispersion interaction in low-dimensional metallic systems}\,\cite{AngyanDispersion2020}. The interaction is much more long-range compared to the textbook results for non-conducting molecules. 
{The model used here describes well a classical plasma, with the electrons free to move in any direction within cylindrical walls. This is directly relevant to a semiconductor cylinder, many atoms thick, and lightly doped so that the mobile electrons are non-degenerate. However, the classical approach used here partly neglects some physical attributes of different elongated molecules such as single-strand conducting polymers, small-radius conducting nanotubes, gold nano-wires, and DNA. These systems are typically a few {\AA}ngstr{\"o}ms's thick. The electron clouds are sometimes degenerate following Fermi statistics}\,\cite{MahanBook}. {However, one should stress that simplified models often reveal the essential physics. As one example Bostr\"om and Sernelius}
\,\cite{MBostromCurrDrag_1999} {used in the past the so-called "plasmon-pole" model to calculate the van der Waals induced current drag between a pair of two-dimensional electron gas sheets. This simplified model gave the same limiting results and power-laws when using the more sophisticated random-phase approximation (RPA)}\,\cite{MBostromCurrDrag_1999}. {Another example is the van der Waals interaction between a pair of two-dimensional electron gas layers, modeled within the RPA, that has the same asymptotic power-law as a pair of thin conducting plasma layers}\,\cite{BostromSerneliusPhysRevB.61.2204}.  {In the same way our alternative description via a plasma model captures much of the essential physics found in past works by Dobson and Ambrosetti}\,\cite{dobson2023mbd+}.

{Here we will summarise previous work on elongated conducting systems. The existing literature mainly covers two basic approaches:
(i) For molecules only a few atoms thick, the conduction electrons are usually degenerate and their motion in the transverse directions is frozen out by quantum confinement. Conduction electron motion then is best treated as strictly one-dimensional, but the contact singularity of the electron-electron Coulomb interaction is smeared by the finite spatial extent of the electronic wavefunctions in the transverse directions. This approach is used, explicitly or implicitly , in Refs}.\,\cite{dobson2023mbd+,AngyanDispersion2020,chang1971van,dobson2006asymptotics}.
{(ii) For samples that are many atoms thick the conducting electrons may be treated as a (usually classical) plasma, free to move in any direction within a cylindrical boundary. This approach is taken in Refs.}\,\cite{Richmond1972,Davies1973}.
{The regions of validity of these approaches overlap to some degree. The emphasis in the present work, including the new material in Sec.}\,\ref{sec4}, {is on approach (ii)}.

In our theoretical derivations, we expand and analyze the work done by B. Davies {\it et al.}\,\cite{Richmond1972,Davies1973}. We will continuously describe what is new compared to past works. We present the theory for dispersion interaction between two thin conducting cylinders. 
We explore the ground and excited state interactions between a pair of elongated conducting molecules. We focus on the ground-state van der Waals interaction where we derive a $1/R^2$ dependency. Similar results have been found by Ninham and co-workers\,\cite{Richmond1972,Davies1973} for 1-dimensional conducting systems. Chang {\it et al.}\,\cite{chang1971van} found asymptotic van der Waals energy for two conducting chains to be $1/\Big(R^2[\ln(2R/a)]^{3/2}\Big)$ and similar type results were also presented by  Davies {\it et al.}\,\cite{Davies1973} when the radius of cylinder is larger than the characteristic Debye length
and recently using different methods for 1D metals by Rubio and co-workers\,\cite{dobson2006asymptotics}. Often, excited state interaction is between individual atoms within the interacting macromolecules\,\cite{Lechelon2022_sciadv.abl5855,AmbrosettiNatCom2022}. However, we predict that if two elongated (cylindrical) conducting molecules are in an overall excited state the same power-law ($1/R^2$) is found as for the corresponding ground state interaction. This can be contrasted with point-like non-conducting molecules where the ground ($1/R^6$) and excited ($1/R^3$) state interaction have different power-laws.

The outline of this work is as follows. We first briefly review the known results for ground state van der Waals interactions between both non-conducting and conducting cylindrical nanoparticles from different groups considering different geometries (including 3-body repulsion). We then review, clearly, the theory originally developed by Ninham and co-workers for ground-state interaction between conducting cylinders. This theory leads us to a new understanding of the excited-state interaction between elongated conducting nanoparticles. Notably, extremely long-ranged excited state interactions are found, and the energy transfer rate is considered. At the end, we discuss future paths we intend to follow to expand the knowledge of excited state interactions between conducting nanoparticles (e.g. DNA).

{We use Gaussian units throughout}.

\section{Past work on van der Waals interaction between cylinders}
We summarize some of the available power-law dependencies for the ground-state van der Waals interaction for cylindrical geometries in Table.\,\ref{SubhojitPowerLawTable}. Here, references to some past work is given.

\begin{table}[h]
\centering
\begin{tabular}{c c c}
  \hline
  \hline
   System    &    Approximations \,   &   Power-laws \\
    \hline
     \makecell{cylinder$||$cylinder \\ (non-conducting)\,\cite{mitchell1973van,langbein1972van,ambrosetti2016wavelike}} & \makecell{Non-retarded \\limit} & $R^{-5}$ \\
    \hline
       \makecell{cylinder$||$cylinder \\ (conducting)\,\cite{Richmond1972,Davies1973}} & \makecell{Non-retarded \\limit,  $a \ll \lambda_{D}$}  & $R^{-2}$\\
    \hline
      & \makecell{Non-retarded \\limit,  $a \gg \lambda_{D}$} &  $R^{-2}[\ln({R/a})]^{-3/2}$\\
     \hline
     \makecell{concentric identical \\ cylinders\,\cite{paresegian2006}} &  \makecell{Non-retarded \\limit}    &  $R^{-2}$\\
     \hline
    \hline
\end{tabular}

\caption{\label{SubhojitPowerLawTable} Asymptotic power-law dependency for van der Waals interaction for different cylindrical systems. $a$ is radius of the cylinders and $\lambda_{D}$ is the characteristic Debye length. `$||$' sign denotes the cylinders are parallel to each other.}
\end{table}
In Table.\,\ref{SubhojitPowerLawTable}, we present two-body (thin cylinders) van der Waals interactions which all have an attractive nature in non-retarded limit. There are  a detailed work done by Richmond\,\cite{richmond1972many,smith1973van} and  co-workers on three body interactions in triangular configuration. Notably, Richmond {\it et al.}\,\cite{richmond1972many} found a repulsive three-body contribution to the van der Waals interaction. 

\section{Interaction between Elongated Conducting Molecules}

Here we consider two conducting cylindrical particles interacting via a conducting medium. For this type of systems, we need to consider real charge and current fluctuations.   
To displace the charge carriers from their equilibrium position, it is necessary to adhere to the continuity equation, which links  charge and current fluctuations as follows,
\begin{equation}
    \frac{\partial \brho}{\partial t } + \nabla \cdot \J = 0
\end{equation}
where $\brho = n e$, $ \J = n e \V$ are the charge and current densities of the charge carriers of charge $e$,  uniform equilibrium density of charge carriers $n_{0}$, and mass $m$.
In this section, we consider a simple hydrodynamic model\,\cite{martinov1971structure}, extensively discussed in electrolyte systems\,\cite{davies1972van}, notably for thin cylinders in a conducting medium\,\cite{Davies1973} etc. Taking into account of possible applicable forces\,(frictional forces, electromagnetic forces, surface forces), the general dynamical behaviour of free charge carriers is determined in the hydrodynamic approximation by the equation,
\begin{equation}
    \frac{\partial \V}{\partial t} + \nu \V + (\V \cdot \bnabla ) \V  = (e/m) \Big[\E + (\V \times \B/c)\Big] - (\bnabla p/nm) 
    \label{eqn:2}
\end{equation}
where $n$ is density of charge carriers, $\nu$ a coefficient describing  friction, $p$ is the pressure. $\E$ and $\B$ are associated electric and magnetic field. {(Note that the dimension of $\nu$ here is 1/s, so that it is not the same as the ordinary kinematic viscosity in hydrodynamics. The dimension of the latter is cm$^2$/s. The quantity $\nu$ above is rather to be regarded as a phenomenological coefficient.)}

We are only concerned with small deviations from the equilibrium parameters\,($n_{0}, \E_{0}, \B_{0}, \rho_{0}$ and $p_{0}$); therefore, we linearize Ninham's model system Eq.\,(\ref{eqn:2}) in Appendix.\,(\ref{appendixa}), resulting an effective equation of motion of charge carriers as,
\begin{equation}
     \frac{\partial \V}{\partial t} + \nu \V = (e/m) \E - (\bnabla p/nm) 
     \label{eqn:3}
\end{equation}
Introducing $\J$ and $\brho$  in place of $\V$ and $n$, Eq.\,(\ref{eqn:3}) can be recast as,
\begin{equation}
    \frac{\partial \J}{\partial t} + \nu \J + s^{2} \bnabla \brho   = ( \omega^{2}_{p}/ 4 \pi) \E
    \label{eqn:4}
\end{equation}
with $\omega_p^2= 4\pi n_0e^2/m$ the squared plasma frequency. 
We substitute $\bnabla p$ with $(\partial p/\partial n) \bnabla n$, where $s^2 = m^{-1} (\partial p/\partial n)$, $s$ represents the isothermal sound velocity of the charge carriers. This approach of Davies {\it et al.} means that the sound velocity appears in the dispersion relations in the cylinders, the propagation velocity of compressional waves playing an important role.

The electric field can be derived from an isotropic dielectric scalar potential $\Phi$ as, $\E = - \bnabla \Phi$. If we take Fourier analysis with respect to time, we have system of equations,
\begin{equation}
    \bnabla \cdot \E = 4 \pi \brho, \label{eqn:5}\\
   \end{equation}
\begin{equation}
  \bnabla \cdot \J = i \omega \brho, \label{eqn:6}  
\end{equation}
\begin{equation}
    (- i \omega + \nu)\,\J + s^{2} \bnabla \brho = - ( \omega^{2}_{p}/ 4 \pi) \bnabla \Phi
    \label{eqn:7}
\end{equation}
It is clearly apparent from these set of equations that all the variables have time dependency $\exp(-i \omega t)$. From Eqs.\,(\ref{eqn:5}) and ({\ref{eqn:6}}), Eq.\,(\ref{eqn:7}) can be reformulated as 
\begin{equation}
       \bnabla^{2} \brho  + (\omega^{2} + i \nu \omega -\omega^{2}_{p})/s^{2}\, \brho= 0
    \label{eqn:8}
\end{equation}

It is a boundary value problem where one boundary condition is that the potential $\Phi$ be continuous at the boundary of the molecules which is equivalently the transverse component of \E. After eliminating $\brho$, one can find a further relation from Eqs.\,(\ref{eqn:5}) and ({\ref{eqn:6}}) as $\bnabla \cdot \D = 0$ where $\D \equiv [ \epsilon \E + (4 \pi i/\omega) \J ]$ is displacement vector. Another boundary condition requires the normal component of $\D$ to be continuous at the molecule's boundary. In this scenario, the intermediate medium is a conducting dielectric region, which means that the normal component of $\J$ must be zero at the surface. Therefore, it is essential to ensure the continuity of $\partial \Phi/\partial n$ at the surface.

Another way of approach in the description of propagating waves in conducting media  would be to allow the wave vector $\bf k$   in the cylinders to be complex, implying that its imaginary part contains the electric conductivity $\sigma$. As is known in ordinary electrodynamics of dielectric media, the complex permittivity $\hat{\epsilon}$  can be expressed as $\hat{\epsilon} = \epsilon + i4\pi \sigma/\omega$, where $\epsilon$  is the real part. Thus
\begin{equation}
    k = \sqrt{\hat{\epsilon}}  \,\omega/c = (n_r\omega/c)\sqrt{1+i4\pi \sigma/(\epsilon \omega)},
\end{equation}
where $n_r=\sqrt{\epsilon}$ is the real refractive index. A very readable account of this formalism can be found in J. A. Stratton, {\it Electromagnetic Theory}\,\cite{stratton1941}, McGraw-Hill, New York, 1941, Ch. IX. 

Whereas in the last equation we kept the permittivity arbitrary, we have so far assumed  that  $\varepsilon =1$ in all regions.

\subsection{General setup of our system}
 \begin{figure}[!h]
    \centering
    \includegraphics[width = 1 \columnwidth]{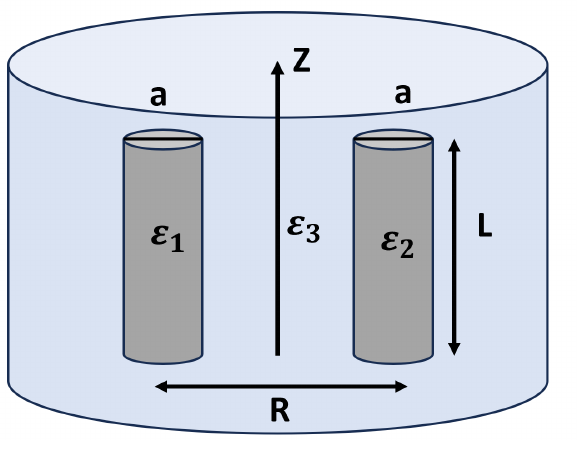}
    \caption{ (Colors online) Schematic representation of two conducting elongated molecules in a media where $\varepsilon_1$ and  $\varepsilon_2$ dielectric functions of the molecules and $\varepsilon_3$ be dielectric function for the media.}
    \label{fig:scheme}
 \end{figure}
For our convenience, we make the assumption that the molecules can be represented as conducting cylinders. In this scenario, we are examining two elongated cylinders depicted in Fig.\,(\ref{fig:scheme}), each with a radius $a$ and length $L$. The first cylinder, characterized by susceptibility $\varepsilon_1$, is positioned parallel to the second cylinder with susceptibility $\varepsilon_2$, and they share a common z-axis aligned with the long axis of the cylinders, separated by a distance $R$. These cylinders are placed in a medium with susceptibility $\varepsilon_3$. The general solution of Eq.\,(\ref{eqn:8}) in Fourier space in cylindrical coordinates can be written as,
\begin{equation}
    \brho\,(r,\theta,z) = \sum_{k}  \int_{-\infty}^{\infty} d\omega\,\brho(r,\theta)  \exp[{-i (k z + \omega t)}] 
    \label{eqn:9}
\end{equation}
and we can extract the general form of potential $\Phi$ from Laplace's equation, $\bnabla^{2} \Phi  = - 4\pi \brho $, which turns to be a similar form like charge density $\brho$ in Eq.\,(\ref{eqn:9}). Now we can easily write down Eq.\,(\ref{eqn:8}) in terms of $\brho(r,\theta)$ as,
\bea
\Big[\bnabla^{2}_{r,\theta} - u^{2}\Big] \brho(r,\theta) = 0, \label{eqn:10}\\
s^{2} u^{2} = (s^{2} k^{2} + \omega^{2}_{p} -\omega^{2} - i\nu\omega) \label{eqn:11}
\eea
where $\bnabla^{2}_{r,\theta}$ is two-dimensional Laplacian in radical polar coordinates. Similarly the equation for  potential $\Phi(r,\theta) $ is 
\begin{equation}
    \Big[\bnabla^{2}_{r,\theta} - u^{2}\Big] \Phi(r, \theta) = -4\pi \brho({r, \theta})
    \label{eqn:12}
    \end{equation}
 While current fluctuations are allowed within the cylinders, they are not allowed in the background medium, it means $\brho = 0$. For simplicity, as of now we ignore the collisions between the free charge carriers ($\nu = 0$).  This simplifies Eq.\,(\ref{eqn:12}). The position of any arbitrary point can be presented by two different sets of cylindrical coordinates (for cylinder 1, $(r_1, \theta_1, z)$ and for cylinder 2, $(r_2, \theta_2, z)$). 
  We can immediately write down the normalized solutions [extensively discussed in Appendix.\,(\ref{appendixb})] to Eq.\,(\ref{eqn:12}) as,
\begin{subequations}
    \begin{gather}
        (1) \indent    \text{Inside the cylinders} \,\quad 
        \Phi_{\text{in}} \equiv \Phi_{i},\,  (i = 1,2) \label{eqn:13a} \\
        (2) \indent   \text{Outside the cylinders} \,\quad 
        \Phi_{\text{out}} \equiv \Phi_{3} \label{eqn:13b}
    \end{gather}
\end{subequations}
where $i$ denotes the cylinder number and 
\begin{equation}
    \Phi_{i} = \sum_{m} A^{(i)}_{m} \exp({i m \theta_{i})}\Bigg[ \frac{I_{m}(kr_{i}) - \gamma I_{m}(ur_{i})}{I_{m}(ka) - \gamma_{i} I_{m}(ua)}\Bigg]
    \label{eqn:14}
\end{equation}
where $\gamma = k \omega^{2}_{p} I^{\prime}_{m}(ka)/u \omega^{2} I^{\prime}_{m}(ua) $
\begin{equation}
    \Phi_{3} = \sum_{i,m} C^{(i)}_{m} \exp({i m \theta_{i})} \Bigg[ \frac{K_{m}(kr_{i})}{K_{m}(ka)}\Bigg]
    \label{eqn:15}
\end{equation}
where $I_{m}$ and $K_{m}$ are modified first and second kind of Bessel functions in standard notations, and $A_{m}$'s, $C_{m}$'s are coefficients which we need to determine. The potentials are shown in Eqs.\,(\ref{eqn:14}) and (\ref{eqn:15}) conventionally continuous at the boundaries of the cylinders. Since we need to satisfy the normal of components of the displacement vector $\D$ across the cylinder surfaces. To do so, we need to express potential outside the cylinders $\Phi_{3}$ entirely in terms of the coordinates of one cylinder\,[here we consider   radical polar coordinate of first cylinder ($r_1,\theta_{1})$]. From Graf's summation formula\,\cite{watson1958}, we know
\begin{equation}
    K_{m}(kr_{2}) \exp({im\theta_{2}}) = \sum_{n = -\infty}^{\infty} K_{m+n}(kR) I_{n}(kr_{1}) \exp({im\theta_{1}})
    \label{eqn:16}
\end{equation}
Now Eq.\,(\ref{eqn:15}) will look like
\begin{equation}
\begin{aligned}
    \Phi_{3} = \sum_{m = -\infty}^{\infty} \Bigg(C^{(1)}_{m}  \Bigg[ \frac{K_{m}(kr_{1})}{K_{m}(ka)}\Bigg] + \sum_{n = -\infty}^{\infty}   \Bigg[\frac{K_{m+n}(kR)}{K_{m}(ka)}\Bigg]\\ C^{(2)}_{n} I_{n}(kr_{1})  \Bigg)\exp({i m \theta_{1})}
    \end{aligned}
    \label{eqn:17}
\end{equation}
we obtain first relation between the coefficients $C^{(1)}_{m}, C^{(2)}_{n} $ and $A^{(1)}_{m}$ from the continuity of potential $\Phi$ at the boundary of cylinder 1 by comparing Eqs.\,(\ref{eqn:14}) and (\ref{eqn:17}) as
\begin{equation}
   A^{(1)}_{m} =  C^{(1)}_{m} + \sum_{n = -\infty}^{\infty} C^{(2)}_{n}  \Bigg[\frac{K_{m+n}(kR)}{K_{m}(ka)}\Bigg] I_{n}(ka) 
   \label{eqn:18}
\end{equation}
Now if we satisfy the second boundary conditions, the continuity of $\epsilon (\partial \Phi/\partial n)$ across the surface of cylinder 1 ($r_1 = a $) , we will get another relation between the coefficients $C^{(1)}_{m}, C^{(2)}_{n} $ and $A^{(1)}_{m}$ after taking derivative of Eqs.\,(\ref{eqn:14}) and (\ref{eqn:17}) with respect to $r_1$ as follows,
\begin{equation}
\begin{split}
   \epsilon_1 A^{(1)}_{m} \Bigg[ \frac{k I^{\prime}_{m}(ka) - \gamma u I^{\prime}_{m}(ua)}{I_{m}(ka) - \gamma I_{m}(ua)}\Bigg]
   =  \epsilon_3 k \Bigg(C^{(1)}_{m}  \Bigg[ \frac{K^{\prime}_{m}(ka)}{K_{m}(ka)}\Bigg] + \\ \sum_{n = -\infty}^{\infty} C^{(2)}_{n}  \Bigg[\frac{K_{m+n}(kR)}{K_{m}(ka)}\Bigg] I^{\prime}_{n}(ka)  \Bigg)
   \end{split}
   \label{eqn:19}
\end{equation}
These Eqs.\,(\ref{eqn:18}) and (\ref{eqn:19}) can be conveniently expressed in terms of a scattering matrix $\widetilde{M}$ after eliminating $A^{(1)}_{m}$ from both of these equations as, $\widetilde{C} = \widetilde{M} {\widetilde{C}}^{\prime} $  {where $\widetilde{C}$ and $\widetilde{C}'$ are column matrices with dimension $m \times 1$ and $n \times 1$ respectively}. 
\begin{equation}
    \begin{aligned}
  \colorbox{white}{$\widetilde{C} \equiv \Big[\widetilde{C} \Big]_{m \times 1}   =   \begin{pmatrix} C^{(1)}_{1} \\ C^{(1)}_{2} \\ \vdots \\ C^{(1)}_{m} \end{pmatrix}  \quad  \&  \quad \widetilde{C}' \equiv \Big[\widetilde{C}' \Big]_{n \times 1} = \begin{pmatrix} C^{(1)}_{1} \\ C^{(1)}_{2} \\ \vdots \\ C^{(1)}_{n} \end{pmatrix} $}
    \end{aligned}
\end{equation}
{$C^{(1)}_{m}$ and $C^{(1)}_{n}$ are defined in Eq}.\,(\ref{C5}). If we follow the same procedure for cylinder 2 and satisfy the boundary conditions, we find a similar expression, ${\widetilde{C}}^{\prime} = \widetilde{N} \widetilde{C}$, where the matrix elements of scattering matrix $\widetilde{M}$, discussed in Appendix (\ref{appendixc}), are given by 
\begin{equation}
\begin{split}
    M_{mn} = - \frac{\Big[\epsilon_{3} - \epsilon_{1m}(k,\omega)\Big] K_{m+n}(kR)}{\Big[ \epsilon_{3} K^{\prime}_{m}(ka)/K_{m}(ka) -  \epsilon_{1m}(k,\omega) I^{\prime}_{m}(ka)/I_{m}(ka) \Big]} \\ \times \frac{I^{\prime}_{m}(ka)}{K_{m}(ka)}
    \end{split}
    \label{eqn:20}
\end{equation}
where 
\begin{equation}
   \epsilon_{1m}(k,\omega) = \frac{\epsilon_{1}\Big[ 1- \omega^{2}_{p}/\omega^2\Big]}{\Big[1- \gamma I_{m}(ua)/I_{m}(ka)\Big]} 
\end{equation}
for the matrix elements $N_{mn}$, we need to substitute $1\rightarrow 2$ in $M_{mn}$. $\epsilon_{1m}(k,\omega)$ is the effective susceptibility of conducting cylinder. {We used tilde sign to denote the scattering matrices for cylinder 1 and cylinder 2 cases separately as  $\widetilde{M}$ and $\widetilde{N}$ and the matrix elements are presented as $M_{mn}$ and $N_{nm}$. The dimensions are }
\begin{equation}
   \colorbox{white}{$ \widetilde{M} \equiv \Big[\widetilde{M}\Big]_{m\times n} = M_{mn} \quad \& \quad  \widetilde{N} \equiv \Big[\widetilde{N}\Big]_{n\times m} = N_{nm} $}
\end{equation}
for the matrix elements $N_{mn}$, we need to substitute $1\rightarrow 2$ in $M_{mn}$.
Now from the expressions $\widetilde{C} = \widetilde{M} {\widetilde{C}}^{\prime} $ and  ${\widetilde{C}}^{\prime} = \widetilde{N} \widetilde{C}$, we deduce a dispersion relation which determines surface modes as follows,
\begin{equation}
    \mathcal{D}(\omega) \equiv \Det(\I- \widetilde{\Omega}) = 0; \quad \widetilde{\Omega} =  \widetilde{M}\widetilde{N}
    \label{eqn:22}
\end{equation}

Eq.\,(\ref{eqn:22}) provides a complete solution to the problem. 
{The free energy of interaction is formally denoted by $G(R,T)$ in terms of the allowed surface modes $\{\omega_{i}\}$ in the dispersion relation Eq.}\,(\ref{eqn:22}) {can be expressed as}\,\cite{NinhamParsegianWeiss1970,physics6010030}
\begin{equation}
    \colorbox{white}{$ G(R,T) = \frac{1}{2 \pi}\int_{0}^{\infty} dk k [G_{R}(k) - G_{\infty}(k)] $}
    \label{eqn:244}
\end{equation}
{where $G_{R}(k) = - k_{B}T \sum_{i} \ln\Big[{2 \sinh{\Big(\frac{\hbar \omega_{i}(k)}{2k_{B}T}}\Big)\Big]} $. In this expression, the summation encompasses all the real solutions of Eq}.\,(\ref{eqn:22}), {and to compute this summation for $\{\omega_{i}\}$, we consider the identity as after transforming summation over momenta to an integral  as $\sum_{k} \to \frac{1}{2\pi} \int_{0}^{\infty}dk k$}.
\begin{equation}
 \colorbox{white}{$   \sum_{i} g(\omega_i)=\frac{1}{2 \pi i} \oint_C g(\omega) \frac{1}{\mathcal{D}(\omega)}\frac{d}{d\omega}\mathcal{D}(\omega)d\omega $}
\end{equation}
{ where the contour $C$ traverses the imaginary axis in a counterclockwise manner from} $+ i \infty$ to $- i\infty$, {encompassing critical points where} $\mathcal{D}$ {vanishes but excluding the poles of} $g$. {To ensure the applicability of Eq.}\,(\ref{eqn:277}), {both} $\mathcal{D}$ {and} $g$ {need to be analytic. Given that} $\widetilde{g}(\omega)=  \ln \Big[2 \sinh \Big(\frac{\hbar \omega(k)}{2k_{B}T}\Big)\Big]$ {exhibits branch cuts, it is advisable to further develop it as} 
 \begin{equation}
    \widetilde{g}(\omega) = \frac{\beta \hbar \omega}{2}-\sum_{n=1}^{\infty} \frac{1}{n} e^{-n \beta \hbar \omega},\quad \beta = \frac{1}{k_{B}T}
 \end{equation}
{Relative permittivity is used to assess} $\mathcal{D}$ {in the upper half plane, where} $\mathcal{D}(\omega)$ {approaches 1. Consequently,} $G_{R}(k)$ {can be expressed as when} $\omega = i\xi$.
\begin{equation}
\begin{aligned}
   G_{R}(k) =  \frac{\hbar}{2} \sum_{i} \omega_{i} + \frac{\hbar}{2 \pi} \sum_{n=1}^{\infty} \int_{-\infty}^{\infty}\cos(n \beta \hbar \xi)\ln \mathcal{D}(i\xi)d\xi 
- \\ \frac{ \hbar i}{2 \pi} \sum_{n=1}^{\infty} \int_{0}^{\infty}\sin(n \beta \hbar \xi)\ln\bigg[\frac{\mathcal{D}(i\xi)}{\mathcal{D}(-i\xi)}\bigg]d\xi
\end{aligned}
\label{eqn:277}
\end{equation}
{The contribution from the third term of right hand side of Eq.}\,(\ref{eqn:277}) {is identically zero if} $\mathcal{D}(\pm i\xi)$ {is an even function and using the identity} $\sum_{n=1}^{\infty} \cos{nx} = \pi \sum_{n = -\infty}^{\infty} \delta(x- 2\pi n) -\frac{1}{2}$, {we can express Eq.}\,(\ref{eqn:277}) {as}
\begin{equation}
    G_{R}(k) = k_{B}T \sum_{n=0}^{\infty}{}' \ln \mathcal{D}(i\xi_{n})
    \label{eqn:288}
\end{equation}
{where  the Matsubara frequency} $\xi_n=2 \pi k T n/\hbar$. {After a little algebra substituting Eq.}\,(\ref{eqn:288}) {into Eq.}\,(\ref{eqn:244}), {we obtain the final free energy expression as Eq.}\,(\ref{eqn:299}).

\subsection{van der Waals interaction between two elongated (cylindrical) conducting molecules}
\label{vdWCondCyl}

It is a well-known fact that the van der Waals interaction per unit length (for a cylinder with length $L$) can be written as\,\cite{Davies1973,Maha},
\begin{equation}
\begin{split}
    G(a,R,T) \simeq \frac{k_{B}T}{\pi}  {\sum_{n=0}^{\infty}}^{\prime} \int_{0}^{\infty} dk\,\ln \mathcal{D}(i \xi_n)  
     \end{split}
     \label{eqn:299}
\end{equation}
following the 1973 work by Ninham and co-workers\,\cite{Davies1973} this can be simplified for a pair of equal thin conducting rodlike molecules (with radii $a\ll R$),

\begin{equation}
\begin{split}
\label{FreeEnergyApprox}
    G(a,R,T) \simeq - \frac{k_{B}T}{4\pi} a^4 {\sum_{n=0}^{\infty}}^{\prime} \int_{0}^{\infty} dk k^4 K^2_{0}(kR) \\ \Bigg(\frac{1- \epsilon_{10}(k,i\xi_{n})}{1- \frac{1}{2} (ka)^2 \ln({ka})\epsilon_{10}(k, i \xi_{n})}\Bigg)^{2}
     \end{split}
\end{equation}

We compared this numerically with the two limits derived by Davies {\it et al.}\,\cite{Davies1973}. In the appropriate limits the deviations between the non-retarded free energy given by Eq.\,(\ref{FreeEnergyApprox}) and the relevant asympote is less than one percent. First when $a \ll \lambda_{D}$,
\begin{equation}
    E(a,R) \simeq - \frac{\hbar \omega_{p} a^4}{64 \pi \lambda_{D} R^2}
\end{equation}
and secondly when $a \gg \lambda_{D}$

\begin{equation}
    E(a,R) \simeq - \frac{\hbar \omega_{p} a}{8\sqrt{2} \pi  R^2 [\ln({R/a})]^{3/2}}.
\end{equation}
This can be compared with the corresponding interaction energy for a pair of thin non-conducting elongated (cylindrical) molecules which has a $1/R^5$-dependence\,\cite{mitchell1973van,langbein1972van}.   In the case of point-like molecules one finds an even faster decay with a $1/R^6$-dependence\,\cite{london1930theorie}. 
For large distances, it is known that only the zero frequency (entropic) part contributes and this term leads to the following long-range contribution consistent with Davies {\it et al.}\,\cite{Davies1973} using the same arguments derived in Appendix.\,(\ref{appendixd}), 
\begin{equation}
\begin{aligned}
\label{FreeEnergyApprox1 }
    G^{\text{vdW}}_{n=0}(a,R,T) &\simeq - \frac{k_{B}T}{2 \pi} \int_{0}^{\infty} dk M_{00}^2 \\ &\simeq  \frac{-k_{B}T \omega^{4}_{p} a^4}{8 \pi s^4 } \int_{0}^{\infty} dk  \frac{K^{2}_{0}(kR)}{\Big[1- \frac{1}{2} \Big(\frac{a}{\lambda_{D}}\Big)^2 \ln(ka)\Big]^2} \\ &\simeq  \colorbox{white}{ $ - \frac{ \pi k_{B}Ta^2\omega^{2}_{p}}{8 s^2 R [\ln({R/a})]^{2}}  \quad a \gg \lambda_{D}  $}
     \end{aligned}
\end{equation}


We will in the following motivate the above equations. Though it is true that Eq.\,(\ref{eqn:22}) provides the formal solution of our problem as it gives the different surface modes which in turn can be used to deduce the van der Waals interaction but analysis of the determinant is one of the key factors for our calculation. In this section we confine an explicit analysis to a special case of ``thin'' cylinders. 
{When the two cylinders come close together the basic assumption that} $a \ll R$ {breaks down and we also need to consider other contributions (e.g. from bound electron that leads to non-metallic short-ranged contributions and the discrete nature of the surface atoms)}.
   {More carefully, the matrix elements defined in Eq.}\,(\ref{eqn:20}) {can only be expanded in powers of} $ka$ {and} $kR$ {in} $kR \lesssim 1$ {region when the cylinders are far apart. Then approximated matrix elements can be represented as,}
   \begin{equation}
       M_{mn} \simeq \Big(\frac{a}{R}\Big)^{m+n} \frac{(m+n-1)!}{m!(n-1)!} < \Big(\frac{a}{R}\Big)^{m+n} 2^{m+n}
   \end{equation}
{So with increasing} $m$ and $n$, {these matrix elements will decrease rapidly. So we obtain convergent values. On the other hand in region} $k \gg 1/R $, {we can use an asymptotic expansion of the modified Bessel function of second kind} $K_{m+n}(kR)$ {to deduce the matrix elements as}
\begin{equation}
    M_{mn} \simeq \sqrt{\frac{1}{kR}} e^{-kR}
\end{equation}
{This also gives finite contributions with increasing}  $m$ {and} $n$. {So in the long separation and short separation, we obtain finite contribution from the matrix elements but in the different regions.} 
In order to carry out this special case of thin cylinders, we follow the method developed by Ninham and co-workers\,\cite{SmithMitchellNinham1970} in the non-retarded limit between cylinders. In the region $kR \lesssim 1$, we can expand the modified Bessel functions of argument $ka$ in the matrix elements $M_{mn}$ and $N_{mn}$ and only consider leading order terms. For small arguments, we list the behaviour of modified Bessel function as 
\begin{equation}
\begin{aligned}
    K_{0}(r) \sim - \ln r,\quad  K_{m}(r) \sim \frac{1}{2}\Gamma(m)\Big(\frac{1}{2} r\Big)^{-m}, \quad I_{0}(r) \sim 1 \\
    I_{m}(r) \sim \Big(\frac{1}{2} r\Big)^{m}/\Gamma(m+1), \quad I^{\prime}_{0}(r) \sim \frac{r}{2}, \quad K^{\prime}_{0}(r) \sim -\frac{1}{r}
     \end{aligned}
     \label{eqn:24}
\end{equation}
We know that in the non-retarded limit, the main contribution comes from $M_{00}N_{00}$ term. Here we take the cylinders are made of same materials, $\epsilon_{1} = \epsilon_{2}$ and the background medium is considered to be vacuum, $\epsilon_3 = 1$. These assumptions significantly made a simple problem as now $M_{00} = N_{00}$. Now we need to evaluate the determinant in Eq.\,(\ref{eqn:22}), as
\begin{equation}
    \Det(M^2 - \I) = 0 \implies M_{00} = \pm 1
    \label{eqn:25}
\end{equation}
which provides the surface modes right away as,
\begin{equation}
    \omega_{\pm} = sk\sqrt{1  \pm 2 K_{0}(kR)}
    \label{eqn:26}
\end{equation}
In the mode summation method the non-retarded van der Waals energy between two thin conducting molecules can be obtained by taking the separation dependent part of the zero-point energy\,\cite{Maha,Ser2018},
\begin{equation}
     \frac{E^{\text{vdW}}(R)}{L}=\frac{\hbar}{4 \pi} \int dk [\omega_+(R)+\omega_-(R)- \omega_+(\infty)- \omega_-(\infty)]
     \label{eqn:28}
\end{equation}
Due to the smallness of parameter $a/R$ implies that $K_{0}(ka) \gg K_{0}(kR)$, which allows the square roots of Eq.\,(\ref{eqn:26}) to be expanded to second order terms. Now the approximated surface modes are 
\begin{equation}
    \omega_{\pm} \simeq sk\Big(1 \pm K_{0}(kR) - \frac{1}{2}K^{2}_{0}(kR) \Big)
    \label{29}
\end{equation}
The first term is a divergent term (the dispersion  relation for pure sound waves) which will be compensated since it is the same when the two cylinders are infinitely far apart. So the van der Waals interaction between two conducting cylinders is 
\begin{equation}
\begin{aligned}
    \frac{E^{\text{vdW}}(R)}{L} &\simeq - \frac{\hbar s}{4 \pi} \int_{0}^{\infty} dk k K^{2}_{0}(kR) \\
    &\simeq - \frac{\hbar s}{8\pi R^2}
    \end{aligned}
    \label{30}
\end{equation}

\section{Resonance Interaction between  identical Molecules}

\label{sec4}
\subsection{Excited state interaction between non-conducting molecules}

Before considering the resonance interaction between a pair of elongated (cylindrical) conducting molecules in an excited state we first rehearse the theory of two identical non-conducting molecules where one initially is in its ground state and the other is in an excited state.  This state can also be represented by a superposition of states: one symmetric and one antisymmetric with respect to the interchange of the molecules. While the symmetric state is likely to decay into two ground-state molecules, the antisymmetric state can be quite long-lived. The system can thus be trapped in the antisymmetric state\,\cite{Stephen}. The energy migrates back and forth between the two molecules until either the two molecules move apart or a photon is emitted away from the system. First-order dispersion interactions are caused by this coupling of the system, i.e. the energy difference between the two states is separation ($R$) dependent.
In the case of two identical molecules, the resonance condition can be separated into one anti-symmetric and one symmetric part. Since the excited symmetric state has a much shorter lifetime than the anti-symmetric state the system can be trapped in an excited anti-symmetric state. The resonance interaction energy of this antisymmetric state is,
\begin{equation}
\label{Eq5}
E^{\text{res}}(R)= \hbar [\omega_{r} (R)-\omega_{r} (\infty)].
\end{equation}
 The resonance interaction energy is within this approximation, and using the definitions of the oscillator strength and the static polarizability \cite{McLachlan},
 \begin{equation}
E^{\text{res}}(R)=p^2 T(R|\omega_j)\propto 1/R^3,
\label{Eq8}
\end{equation}
where $p$ is the magnitude of the transition dipole moment. This is the correct textbook result for low-temperature resonance interaction energy between a pair of non-conducting molecules in an excited state in the non-retarded limit.  This result comes out identical from fundamentally different approaches.

\subsection{Excited state interaction between elongated (cylindrical) conducting molecules}

The resonance interaction energy between two thin elongated molecules is found by assuming that the antisymmetric mode is much more long lived than the symmetric mode,

\begin{equation}
   \frac{E^{\text{res}}(R)}{L}=\frac{\hbar}{2 \pi} \int dk [\omega_{-}(R)- \omega_{-}(\infty)]
\end{equation}
If we follow the same procedure as discussed in Sec.\,\ref{vdWCondCyl}, we predict that the resonance interaction goes as,
\begin{equation}
   \frac{E^{\text{res}}(R)}{L}\approx - \frac{5\, \hbar s}{8\pi R^2}
   \label{ResNonRet}
\end{equation}
Following in the footsteps of past work one can also write the solution to the resonance interaction (when $a\lesssim R$) when antisymmetric or symmetric mode is excited, 
\,\cite{Bostrom1}
\begin{equation}
\begin{split}
    G^{\text{res}}(a,R,T) \simeq  \frac{\pm 2 k_{B}T}{\pi}  {\sum_{n=0}^{\infty}}^{\prime} \int_{0}^{\infty} dk\,\ln \Det(\I- \widetilde{M})
     \end{split}
\end{equation}

\begin{equation}
\begin{split}
    G^{\text{res}}(a,R,T)\simeq   \frac{\mp2 k_{B}T}{\pi}  {\sum_{n=0}^{\infty}}^{\prime} \int_{0}^{\infty} dk M_{00}  
     \end{split}
     \label{FreeResonance}
\end{equation}
From the zero frequency term in Eq.\,(\ref{FreeResonance}) we derive the entropic long-range resonance interaction between a pair of elongated conducting molecules which is given,

\begin{equation}
\begin{aligned}
    G^{\text{res}}_{n=0}(a,R,T) &\simeq \frac{\mp k_{B}T}{\pi} \int_{0}^{\infty} dk M_{00} \\
   & \simeq \colorbox{white}{$ \mp \frac{ k_{B}T }{2 R \ln\Big(\frac{R}{a}\Big)} \quad a \gg \lambda_{D} $}
     \end{aligned}
     \label{FreeResonancezerofreq}
\end{equation}

In principle, it is possible to take the zero temperature limit of Eq.\,(\ref{FreeResonance}) and receive the $1/R^2$ dependency when  $a \ll \lambda_{D}$  found earlier in Eq.\,(\ref{ResNonRet}).  To take the zero temperature limit we in a standard manner replace the summation over discrete frequencies with an integration over imaginary frequencies by a summation over discrete frequencies \cite{Dzya,Maha}.

\begin{equation}
k_B T \sum_{n=0}^{\infty}{}',\,\xi_n= \frac{2 \pi k_B T n}{\hbar }\to {{\hbar} \over {2 \pi}} \int_0^\infty d \xi , 
\label{EqC4_34}
\end{equation}
where $k_B$ is the Boltzmann constant and the prime indicates that the
$n=0$ term should be divided by 2.  Now if we consider Eq.\,(\ref{EqC4_34}) and substitute back into Eq.\,(\ref{FreeResonance}) taking into account $R \gg a$, we get
\begin{equation}
\begin{aligned}
     G^{\text{res}}(a,R)\simeq  \pm   \frac{\hbar}{2 \pi^2}  \int_{0}^{\infty} d \xi \int_{0}^{\infty} dk (ka)^2K_{0}(kR)\\
      \Bigg(\frac{\frac{\omega^{2}_{p}}{\xi^2 + k^2 s^2}}{1- \frac{1}{2} (ka)^2 \ln({ka})\Big(1 + \frac{\omega^{2}_{p}}{\xi^2 + k^2 s^2}\Big)}\Bigg)
     \end{aligned}
\end{equation}
We can first carry out  the $\xi$ integration easily and  can drop  $(ka)^2 \ln({ka})$ term as $k \lesssim R^{-1}$, we obtain
\begin{equation}
\begin{aligned}
 G^{\text{res}}(a,R) &\approx \mp \,\frac{\hbar \omega^{2}_{p} a^2 }{4 s \pi}  \int_{0}^{\infty} dk k^2 \frac{K_{0}(kR)}{\Big[1- \frac{1}{2}\, k^2 \Big(\frac{a}{\lambda_{D}}\Big)^2 \ln(ka)\Big]^{\frac{1}{2}}}\\
 &\approx  \colorbox{white}{$ \mp \frac{\hbar \omega_{p} a }{4 \sqrt{2} \pi R^2[\ln(R/a)]^{\frac{1}{2}}} \quad a \gg \lambda_{D} $}
 \end{aligned}
\end{equation}

\subsection{F\"{o}rster Energy Transfer between elongated conducting molecules}

Since we now have an expression for the resonance interaction energy between two identical elongated conducting molecules in an excited configuration we can estimate the F\"{o}rster energy transfer rate. In the strong-coupling limit one defines the rate of "fast" transfer between two identical molecules, one in the ground state and the other in an excited state, as \cite{Forster}:
\begin{equation} 
n \approx 2  |E^{\text{res}}|/(\pi \hbar)\propto 1/R^2,
\label{Eq_fastForster}
\end{equation}
where $E^{res}$ is the resonance energy and $\hbar$ is Planck's constant. F\"{o}rster demonstrated how the transfer rate of both strongly and weakly coupled molecules can be treated within the same formalism \cite{Forster}. Between two weakly interacting molecules, that may in general be different, there is enough that there is an overlap of the energy-bands to have energy transfer. Application of time-dependent perturbation theory gives the following approximation (Fermi golden rule rate) for this "slow" transfer rate\cite{Craig,Forster}:
\begin{equation}
n \approx 2\pi |E^{\text{res}}|^2 \delta/\hbar\propto 1/R^4,
\label{Eq_slowForster}
\end{equation}
where $\delta$ is the "density of final states"  (related to the spread in the energy of the optical band associated with slow energy transfer \cite{Forster}).
Hence, the energy transfer rate between a pair of elongated (cylindrical) conducting molecules, when $a \ll \lambda_{D}$, goes like either $1/R^2$ or $1/R^4$
in the non-retarded limit. In the opposite limit, with $a \gg \lambda_{D}$, a slightly modified energy transfer rate is expected (i.e., $R^{-2} [\ln({R/a})]^{-3/2}$ or $R^{-4} [\ln({R/a})]^{-3}$).
What is most noteworthy we predict from Eq.\,(\ref{FreeResonancezerofreq}) the F\"orster transfer rate goes as $n\propto k_B\,T/R$  (in the fast transfer case) or as $n \propto k_B^2\,T^2/R^2$  (in the "slow" transfer case) in the large distance/high-temperature limit.

\section{Future outlooks}

Our work is obviously relevant for the interaction between conducting polymers, DNA, and other linear molecules. The interaction is as we have discussed peculiar, very long range and the interaction between elongated conducting molecules is strictly non additive. The theory was derived in the 1970s by Ninham and co-workers\,\cite{Maha}. It is also in the more recent book by Ninham and Lo Nostro\,\cite{Ninhb} and in Parsegian's book\,\cite{Pars}. We have explored these topics in some detail in the current work. Ultimately, the question arises if long-ranged interaction can exist between a ground state and an excited state conducting molecule. Clearly, metastable states exist with long-lived eigenvalues separated from the continuum band of linear molecules\,\cite{NinhamNossalZwanzig1969} producing interaction between excited-ground states of cylindrical molecules. There is the potential relation to energy transfer in the pheremone problem\,\cite{Ninhb} and others involving very long-range interactions and recognition for DNA interactions. This is a fundamentally untouched area of research that we aim to explore further. In the longer term, circular dichroism and light-induced electron transfer can be considered. The interaction between a molecule and a cylinder could lead to new models for ground and excited state RNA-DNA interactions. The interaction between conducting cylinders, as models for DNA, in salt where the conduction due to fluctuations of counterions around the cylinders\,\cite{MathiasDavidBarryDNA2002} is of much interest as a model of DNA recognition. Finally, interactions between conducting molecules at or between metals (dielectrics) has a clear connection  with catalysis (e.g., cracking oil in zeolites). Further background to this highly exciting field are outlined in the book "The Language of Shape"\,\cite{HydeLanuage}.

\section*{Conclusions}
\vspace{-0.07in}
Considering the vital importance of the F\"{o}rster energy transfer in biophysics it is important that the underlying theory is done correctly. We have demonstrated an extraordinarily slowly decay rate for both the van der Waals interaction and the
resonance energy interaction for elongated cylindrical conducting molecules. This indicates that for conducting molecules the energy transfer rates are much stronger, and of longer range, than what has been previously assumed. The past results for excited state resonance interaction, discussed for instance articles in Science Advances\,\cite{Lechelon2022_sciadv.abl5855}  and  Nature Communications\,\cite{AmbrosettiNatCom2022}, have thus a limited relevance for a pair of cylindrical conducting molecules in a coupled excited state.

\appendix  
\section{General method to linearize Eq.\,(\ref{eqn:2}) representing a simple hydrodynamic model}
\label{appendixa}
For linearization of Eq.\,(\ref{eqn:2}), we set our parameters as,
\begin{equation}
\begin{split}
n = n_{0} + n_{1} \\
p = p_{0} + p_{1} \\
\rho = \rho_{0} + \rho_{1}\\
\E = \E_{0} + \E_{1}\\
\B  = \B_{0} +\B_{1}
\end{split}
\label{A1}
\end{equation}
by general set up, we consider $ n_{0} \gg n_{1}$,  $ p_{0} \gg p_{1}$, $ \rho_{0} \gg \rho_{1}$, $\E_{0} \gg \E_{1}$ and $\B_{0} \gg \B_{1}$. To remove $p$ from Eq.\,(\ref{eqn:2}), we utilize the information that $p \equiv p(n)$ at a constant temperature $T$, thus we can express it is as a functional derivative,
\begin{equation}
    (nm)^{-1} \bnabla p = (nm)^{-1} (\partial p/ \partial n) \bnabla n
    \label{A2}
\end{equation}
For simplicity we set $\B_{0} = 0 $ and omit the convective term on the LHS of Eq.\,(\ref{eqn:2}), we get
\begin{equation}
     \frac{\partial \V}{\partial t} + \nu \V  = (e/m) (\E_{0} + \E_{1})- (nm)^{-1} (\partial p/ \partial n) \bnabla (n_{0} + n_{1})
     \label{A3}
\end{equation}

Now take Taylor series expansion of pressure $p$ as a function of $n$, we get
\begin{equation}
    p(n) = p(n_{0}) + n_{1} \Big(\partial p/\partial n\Big)_{0} + \bigO (n_{1})^{2}
    \label{A4}
\end{equation}
Now one needs to extract the first order contribution of $n_{1}$ using,
\begin{equation}
\begin{split}
    (\partial p/ \partial n) \bnabla n \simeq (\partial p/ \partial n)_{0} \bnabla n_{0} + (\partial p/ \partial n) \bnabla n_{1} + \\ n_{1} (\partial^{2} p/ \partial n^{2})_{0} \bnabla n_{0}
\end{split}
\label{A5}
\end{equation}
As $n_{0} \gg n_{1}$, $1/n$ can be approximated as $(1/n_{0} - n_{1}/n^{2}_{0})$. If we  use this approximation along with Eq.\,(\ref{A5}) into Eq.\,(\ref{A3}), after some careful algebra it yields a set of equations as,
\begin{equation}
    n_{0} e \E_{0} = (\partial p/ \partial n)_{0} \bnabla n_{0}
    \label{A6}
\end{equation}
\begin{equation}
\begin{split}
    \frac{\partial \V}{\partial t} + \nu \V = (e/m) \E_{1} - (n_{0}m)^{-1} (\partial p/ \partial n)_{0} \bnabla n_{1} \\ - (\bnabla n_{0} /n_{0}m) \Big[n^{-1}_{0} (\partial p/ \partial n)_{0}  + (\partial^{2} p/ \partial n^{2})_{0} \Big] n_{1} 
    \end{split}
    \label{A7}
\end{equation}
It is worth to mention $\E_{0} = 0 $ due to uniform equilibrium density of charge carriers, $\bnabla n_{0} = 0 $. Hence we can drop the last two terms of Eq.\,(\ref{A7}). We can consider $(\partial p/ \partial n)_{0} $ as $kT$ for electrolytes\,[Berry's paper]

\section{Derivation: the Laplacian}
\label{appendixb}
Here we will show how to treat our general solutions more systematically. 
In cylindrical coordinates $(r, \theta, z)$, Laplacian is 
\begin{equation}
    \bnabla^{2} = \overbrace{\frac{1}{r} \frac{\partial}{\partial r}\Big[r\frac{\partial}{\partial r}\Big] + \frac{1}{r^2} \frac{\partial^2}{\partial \theta^2} }^{ \bnabla^{2}_{r,\theta}}+ \frac{\partial^2}{\partial z^{2}}
    \label{B1}
\end{equation}
where $ \bnabla^{2}_{r,\theta}$ is two-dimensional Laplacian in radical polar coordinates. Now if we apply this operator into Eq.\,(\ref{eqn:9}), it immediately give Eq.\,(\ref{eqn:10}) for $\brho(r,\theta)$ and Eq.\,(\ref{eqn:12}) for $\Phi(r,\theta)$. To solve Eq.\,(\ref{eqn:12}), we follow separation of variable approach. Let's take ansatz $\Phi(r,\theta) = R(r)\Theta(\theta)$. If we put this ansatz into Eq.\,(\ref{eqn:12}), it yields
a pair of differential equations as
\begin{subequations}
    \begin{gather}
         \frac{\partial^2 R}{\partial x^2} + \frac{1}{x} \frac{\partial R}{\partial x} - (1 + m^2/x^2)R = 0, \quad x = kr \label{B2a} \\
          \frac{\partial^2 \Theta}{\partial \theta} + m^2 \Theta = 0 \label{B2b} 
    \end{gather}
\end{subequations}
where Eq.\,(\ref{B2a}) is a differential equation for modified Bessel's function. Hence full normalized solution of Eq.\,(\ref{eqn:12}) can be written as,
\begin{equation}
    \Phi_{\text{out}} = \sum_{m = -\infty}^{\infty} C_{m} \exp({im\theta}) \Bigg[ \frac{K_{m}(kr)}{K_{m}(ka)}\Bigg]; \quad  r > a
    \label{B3}
\end{equation}
Here the reason for choosing of $K_m(kr)$ is that this function goes to zero at infinity, and for inside the cylinders $ r < a$ [derivation is given in 1973 paper]
\begin{equation}
    \Phi_{\text{in}} =\sum_{m = -\infty}^{\infty} A_{m} \exp({i m \theta)}\Bigg[ \frac{I_{m}(kr) - \gamma I_{m}(ur)}{I_{m}(ka) - \gamma I_{m}(ua)}\Bigg]
    \label{B4}
\end{equation}
where $\gamma = k \omega^{2}_{p} I^{\prime}_{m}(ka)/u \omega^{2} I^{\prime}_{m}(ua) $. 
\section{Derivation of scattering matrix}
\label{appendixc}
The starting point of this calculation is the second boundary condition across the surface of the cylinder 1. The continuity equation is 
\begin{equation}
    \epsilon_{1} \frac{\partial \Phi_{1}}{\partial r_1}\Big|_{r_1 =a} = \epsilon_{3} \frac{\partial \Phi_{3}}{\partial r_1}\Big|_{r_1 =a}
    \label{C1}
\end{equation}
After omitting $\exp({im\theta_{1}})$, we obtain a connection between the coefficients $C^{(1)}_{m}, C^{(2)}_{m} $ and $A^{(1)}_{m}$from Eqs.\,(\ref{eqn:14}) and (\ref{eqn:17}) as,

\begin{equation}
\begin{split}
   \epsilon_1 A^{(1)}_{m} \Bigg[ \frac{k I^{\prime}_{m}(ka) - \gamma u I^{\prime}_{m}(ua)}{I_{m}(ka) - \gamma I_{m}(ua)}\Bigg]
   =  \epsilon_3 k \Bigg(C^{(1)}_{m}  \Bigg[ \frac{K^{\prime}_{m}(ka)}{K_{m}(ka)}\Bigg] + \\ \sum_{n = -\infty}^{\infty} C^{(2)}_{m}  \Bigg[\frac{K_{m+n}(kR)}{K_{m}(ka)}\Bigg] I^{\prime}_{n}(ka)  \Bigg)
   \end{split}
   \label{C2}
\end{equation}
Now if we compare Eqs.(\ref{eqn:18}) and (\ref{C2}), we got a complicated expression between $C^{(1)}_{m}$ and $C^{(2)}_{n}$ as,
\begin{equation}
    \begin{split}
  C^{(1)}_{m} = - \sum_{n = -\infty}^{\infty} \Biggl[\frac{k \epsilon_{3}- \epsilon_{1} \frac{I_{m}(ka)}{I^{\prime}_{m}(ka)} \Big(\frac{k \frac{I^{\prime}_{m}(ka)}{I_{m}(ka)} - \gamma u \frac{I^{\prime}_{m}(ua)}{I_{m}(ka)}}{1 - \gamma \frac{I_{m}(ua)}{I_{m}(ka)}}\Big)}{k \epsilon_{3}\frac{K^{\prime}_{m}(ka)}{K_{m}(ka)} -  \epsilon_{1} \Big(\frac{k \frac{I^{\prime}_{m}(ka)}{I_{m}(ka)} - \gamma u \frac{I^{\prime}_{m}(ua)}{I_{m}(ka)}}{1 - \gamma \frac{I_{m}(ua)}{I_{m}(ka)}}\Big)}\Biggr] \\
  \times \frac{K_{m+n}(kR)}{K_{m}(ka)} I^{\prime}_{m}(ka) C^{(2)}_{n}
    \end{split}
    \label{C3}
\end{equation}
Plugging the values of $\gamma$ and $u$ into the expression $\Big(k I^{\prime}_{m}(ka)/I_{m}(ka) - \gamma u I^{\prime}_{m}(ua)/I_{m}(ka)\Big)$, we obtain
\begin{equation}
   \Big(k \frac{I^{\prime}_{m}(ka)}{I_{m}(ka)} - \gamma u \frac{I^{\prime}_{m}(ua)}{I_{m}(ka)}\Big) = k \frac{I^{\prime}_{m}(ka)}{I_{m}(ka)} \Big(1- \frac{\omega^{2}_{p}}{\omega^2}\Big)
   \label{C4}
\end{equation}
If we put Eq.\,(\ref{C4}) into Eq.\,(\ref{C3}), it yields 
\begin{equation}
\begin{aligned}
  C^{(1)}_{m} = - \sum_{n = -\infty}^{\infty}    \underbrace{\frac{\Big[\epsilon_{3} - \epsilon_{1m}(k,\omega)\Big] K_{m+n}(kR) \frac{I^{\prime}_{m}(ka)}{K_{m}(ka)}}{\Big[ \epsilon_{3}\frac{K^{\prime}_{m}(ka)}{K_{m}(ka)} -  \epsilon_{1m}(k,\omega) \frac{I^{\prime}_{m}(ka)}{I_{m}(ka)}\Big]}}_{M_{mn}}  \\  \times C^{(2)}_{n}
  \end{aligned}
  \label{C5}
\end{equation}
where $ \epsilon_{1m}(k,\omega) = \frac{\epsilon_{1}\Big[ 1- \omega^{2}_{p}/\omega^2\Big]}{\Big[1- \gamma I_{m}(ua)/I_{m}(ka)\Big]} $. Hence $\widetilde{C} = \widetilde{M} {\widetilde{C}}^{\prime} $. \\

\section{Derivation of zero frequency resonance energy}
\label{appendixd}
{If we consider thin cylinder approximation as $R \gg a$ i.e., we need to drop the term $k^2 a^2 \ln(ka)$ everywhere in the denominator of $M_{00}$ in this calculation.}
\begin{equation}
\begin{aligned}
  G^{\text{res}}_{n=0}(a,R,T)  &  \simeq \frac{\mp k_{B}T}{\pi} \int_{0}^{\infty} dk M_{00} \\
    & \simeq \frac{\mp k_{B}T \omega^{2}_{p} a^2}{2 \pi s^2} \int_{0}^{\infty} dk \frac{K_{0}(kR)}{\Big[1- \frac{1}{2} \Big(\frac{a}{\lambda_{D}}\Big)^2 \ln(ka)\Big]}
    \end{aligned}
     \label{FreeResonancezerofreq}
\end{equation}
This integration can be solved in two different limits
\begin{enumerate}
    \item when $a \ll \lambda_{D}$
    \begin{equation}
    \begin{aligned}
        G^{\text{res}}_{n=0}(a,R,T) &\simeq \mp \frac{ k_{B}T \omega^{2}_{p} a^2}{2 \pi s^2}\int_{0}^{\infty} dk K_{0}(kR) \\
        & \simeq  \mp \frac{ k_{B}T \omega^{2}_{p} a^2}{4 s^2 R}\,\Bigg[1 + \mathcal{O}\Big(\frac{a}{\lambda_{D}}\Big)^2 \Bigg] 
        \end{aligned}
        \end{equation}
    \item when $a \gg \lambda_{D}$. In this limit the maximum contribution will come from the range where $kR \lesssim 1$. we define a smallness parameter $\alpha =a/R \ll 1$. Let $x = k a$ and $k R=x  R/a= x /\alpha$. This leads to ($k=x/a$ and $dk=dx/a$),
    \begin{equation}
        \begin{aligned}
          G^{\text{res}}_{n=0}(a,R,T) &\simeq \mp    \frac{ k_{B}T \omega^{2}_{p} a^2}{2 a \pi s^2}\int_{0}^{\infty} dx \frac{K_{0}(\frac{x}{\alpha})}{\Big[1- \frac{1}{2} \Big(\frac{a}{\lambda_{D}}\Big)^2 \ln(x)\Big]}\\
          & \simeq \mp \frac{ k_{B}T }{\pi a }\int_{0}^{\infty} dx \frac{K_{0}(\frac{x}{\alpha})}{\big[-\ln(x)\big]}
        \end{aligned}
        \label{eqn:74}
    \end{equation}
    As $kR \lesssim 1$, it gives $x \sim \frac{a}{R}$. Since the denominator of Eq.\,(\ref{eqn:74}) can be written as $ -\ln x \sim \ln\Big(\frac{R}{a}\Big) $

\begin{equation}
\begin{aligned}
    G^{\text{res}}_{n=0}(a,R,T)& \simeq \mp \frac{ k_{B}T }{\pi a \ln\Big(\frac{R}{a}\Big)}\int_{0}^{\infty} dx K_{0}\Big(\frac{x}{\alpha}\Big) \\
    & \simeq \mp \frac{ k_{B}T }{2 R \ln\Big(\frac{R}{a}\Big)}\,\Bigg[1 + \mathcal{O}\Big(\frac{\ln R}{a}\Big)^{-1} + \mathcal{O}\Big(\frac{a}{\lambda_{D}}\Big)^2 \Bigg] 
\end{aligned}
\end{equation}
\end{enumerate}





\section*{Author Contributions}
 Bostr\"om designed the study. The analytical modelling was done by Pal supported by Brevik and Bostr\"om.  All authors contributed to the writing and overall analysis of the manuscript.  All authors have approved the final version of the manuscript.

\section*{Conflicts of interest}
There are no conflicts of interest to declare.

\section*{Acknowledgements}
 This research is part of the project No. 2022/47/P/ST3/01236 co-funded by the National Science Centre and the European Union's Horizon 2020 research and innovation programme under the Marie Sk{\l}odowska-Curie grant agreement No. 945339. 
Institutional and infrastructural support for the ENSEMBLE3 Centre of Excellence was provided through the ENSEMBLE3 project (MAB/2020/14) delivered within the Foundation for Polish Science International Research Agenda Programme and co-financed by the European Regional Development Fund and the Horizon 2020 Teaming for Excellence initiative (Grant Agreement No. 857543), as well as the Ministry of Education and Science initiative “Support for Centres of Excellence in Poland under Horizon 2020” (MEiN/2023/DIR/3797).

\providecommand*{\mcitethebibliography}{\thebibliography}
\csname @ifundefined\endcsname{endmcitethebibliography}
{\let\endmcitethebibliography\endthebibliography}{}


\begin{mcitethebibliography}{46}
\providecommand*{\natexlab}[1]{#1}
\providecommand*{\mciteSetBstSublistMode}[1]{}
\providecommand*{\mciteSetBstMaxWidthForm}[2]{}
\providecommand*{\mciteBstWouldAddEndPuncttrue}
  {\def\EndOfBibitem{\unskip.}}
\providecommand*{\mciteBstWouldAddEndPunctfalse}
  {\let\EndOfBibitem\relax}
\providecommand*{\mciteSetBstMidEndSepPunct}[3]{}
\providecommand*{\mciteSetBstSublistLabelBeginEnd}[3]{}
\providecommand*{\EndOfBibitem}{}
\mciteSetBstSublistMode{f}
\mciteSetBstMaxWidthForm{subitem}
{(\emph{\alph{mcitesubitemcount}})}
\mciteSetBstSublistLabelBeginEnd{\mcitemaxwidthsubitemform\space}
{\relax}{\relax}

\bibitem[Bordag \emph{et~al.}(2009)Bordag, Klimchitskaya, Mohideen, and Mostepanenko]{Bordagbook}
M.~Bordag, G.~L. Klimchitskaya, U.~Mohideen and V.~M. Mostepanenko, \emph{{Advances in the Casimir Effect}}, Oxford Science Publications, Oxford, 2009\relax
\mciteBstWouldAddEndPuncttrue
\mciteSetBstMidEndSepPunct{\mcitedefaultmidpunct}
{\mcitedefaultendpunct}{\mcitedefaultseppunct}\relax
\EndOfBibitem
\bibitem[Ninham and Lo~Nostro(2010)]{Ninhb}
B.~W. Ninham and P.~Lo~Nostro, \emph{Molecular Forces and Self Assembly in Colloid, Nano Sciences and Biology}, Cambridge University Press, Cambridge, 2010\relax
\mciteBstWouldAddEndPuncttrue
\mciteSetBstMidEndSepPunct{\mcitedefaultmidpunct}
{\mcitedefaultendpunct}{\mcitedefaultseppunct}\relax
\EndOfBibitem
\bibitem[Sernelius(2018)]{Ser2018}
B.~E. Sernelius, \emph{{Fundamentals of van der Waals and Casimir Interactions}}, Springer International Publishing, 2018\relax
\mciteBstWouldAddEndPuncttrue
\mciteSetBstMidEndSepPunct{\mcitedefaultmidpunct}
{\mcitedefaultendpunct}{\mcitedefaultseppunct}\relax
\EndOfBibitem
\bibitem[London(1930)]{london1930theorie}
F.~London, \emph{Zeitschrift f{\"u}r Physik}, 1930, \textbf{63}, 245--279\relax
\mciteBstWouldAddEndPuncttrue
\mciteSetBstMidEndSepPunct{\mcitedefaultmidpunct}
{\mcitedefaultendpunct}{\mcitedefaultseppunct}\relax
\EndOfBibitem
\bibitem[Cario and Franck(1923)]{Cario}
G.~Cario and J.~Franck, \emph{Z. Physik}, 1923, \textbf{17}, 202–212\relax
\mciteBstWouldAddEndPuncttrue
\mciteSetBstMidEndSepPunct{\mcitedefaultmidpunct}
{\mcitedefaultendpunct}{\mcitedefaultseppunct}\relax
\EndOfBibitem
\bibitem[F\"{o}rster(1965)]{Forster}
T.~i. O. S.~e. F\"{o}rster, \emph{{Modern Quantum Chemistry, Pt. 3}}, Academic, New York, 1965\relax
\mciteBstWouldAddEndPuncttrue
\mciteSetBstMidEndSepPunct{\mcitedefaultmidpunct}
{\mcitedefaultendpunct}{\mcitedefaultseppunct}\relax
\EndOfBibitem
\bibitem[{van Grondelle} \emph{et~al.}(1994){van Grondelle}, Dekker, Gillbro, and Sundstr\"{o}m]{Groendelle1}
R.~{van Grondelle}, J.~P. Dekker, T.~Gillbro and V.~Sundstr\"{o}m, \emph{Biochimica et Biophysica Acta (BBA) - Bioenergetics}, 1994, \textbf{1187}, 1--65\relax
\mciteBstWouldAddEndPuncttrue
\mciteSetBstMidEndSepPunct{\mcitedefaultmidpunct}
{\mcitedefaultendpunct}{\mcitedefaultseppunct}\relax
\EndOfBibitem
\bibitem[Brennen \emph{et~al.}(2000)Brennen, Deutsch, and Jessen]{Brennen}
G.~K. Brennen, I.~H. Deutsch and P.~S. Jessen, \emph{Phys. Rev. A}, 2000, \textbf{61}, 062309\relax
\mciteBstWouldAddEndPuncttrue
\mciteSetBstMidEndSepPunct{\mcitedefaultmidpunct}
{\mcitedefaultendpunct}{\mcitedefaultseppunct}\relax
\EndOfBibitem
\bibitem[Perina(2001)]{Perina}
J.~Perina, \emph{{Coherence and Statistics of Photons and Atoms}}, Wiley, New York, 2001\relax
\mciteBstWouldAddEndPuncttrue
\mciteSetBstMidEndSepPunct{\mcitedefaultmidpunct}
{\mcitedefaultendpunct}{\mcitedefaultseppunct}\relax
\EndOfBibitem
\bibitem[Jones \emph{et~al.}(1996)Jones, Julienne, Lett, Phillips, Tiesinga, and Williams]{Jones}
K.~M. Jones, P.~S. Julienne, P.~D. Lett, W.~D. Phillips, E.~Tiesinga and C.~J. Williams, \emph{Europhys. Lett.}, 1996, \textbf{35}, 85\relax
\mciteBstWouldAddEndPuncttrue
\mciteSetBstMidEndSepPunct{\mcitedefaultmidpunct}
{\mcitedefaultendpunct}{\mcitedefaultseppunct}\relax
\EndOfBibitem
\bibitem[Stephen(1964)]{Stephen}
M.~J. Stephen, \emph{J. Chem. Phys.}, 1964, \textbf{40}, 669\relax
\mciteBstWouldAddEndPuncttrue
\mciteSetBstMidEndSepPunct{\mcitedefaultmidpunct}
{\mcitedefaultendpunct}{\mcitedefaultseppunct}\relax
\EndOfBibitem
\bibitem[McLachlan(1964)]{McLachlan}
A.~McLachlan, \emph{Molecular Physics}, 1964, \textbf{8}, 409--423\relax
\mciteBstWouldAddEndPuncttrue
\mciteSetBstMidEndSepPunct{\mcitedefaultmidpunct}
{\mcitedefaultendpunct}{\mcitedefaultseppunct}\relax
\EndOfBibitem
\bibitem[McLone and Power(1964)]{Mclone}
R.~R. McLone and E.~A. Power, \emph{Mathematika}, 1964, \textbf{11}, 91--94\relax
\mciteBstWouldAddEndPuncttrue
\mciteSetBstMidEndSepPunct{\mcitedefaultmidpunct}
{\mcitedefaultendpunct}{\mcitedefaultseppunct}\relax
\EndOfBibitem
\bibitem[Bostr\"{o}m \emph{et~al.}(2003)Bostr\"{o}m, Longdell, Mitchell, and Ninham]{Bostrom1}
M.~Bostr\"{o}m, J.~Longdell, D.~Mitchell and B.~Ninham, \emph{Eur. Phys. J. D}, 2003, \textbf{22}, 47--52\relax
\mciteBstWouldAddEndPuncttrue
\mciteSetBstMidEndSepPunct{\mcitedefaultmidpunct}
{\mcitedefaultendpunct}{\mcitedefaultseppunct}\relax
\EndOfBibitem
\bibitem[Dobson and Ambrosetti(2023)]{dobson2023mbd+}
J.~F. Dobson and A.~Ambrosetti, \emph{J. Chem. Theory and Computation.}, 2023, \textbf{19}, 6434--6451\relax
\mciteBstWouldAddEndPuncttrue
\mciteSetBstMidEndSepPunct{\mcitedefaultmidpunct}
{\mcitedefaultendpunct}{\mcitedefaultseppunct}\relax
\EndOfBibitem
\bibitem[Mitchell \emph{et~al.}(1973)Mitchell, Ninham, and Richmond]{mitchell1973van}
D.~Mitchell, B.~Ninham and P.~Richmond, \emph{Biophys. J.}, 1973, \textbf{13}, 359--369\relax
\mciteBstWouldAddEndPuncttrue
\mciteSetBstMidEndSepPunct{\mcitedefaultmidpunct}
{\mcitedefaultendpunct}{\mcitedefaultseppunct}\relax
\EndOfBibitem
\bibitem[Langbein(1972)]{langbein1972van}
D.~Langbein, \emph{Physik der kondensierten Materie}, 1972, \textbf{15}, 61--86\relax
\mciteBstWouldAddEndPuncttrue
\mciteSetBstMidEndSepPunct{\mcitedefaultmidpunct}
{\mcitedefaultendpunct}{\mcitedefaultseppunct}\relax
\EndOfBibitem
\bibitem[Ambrosetti \emph{et~al.}(2016)Ambrosetti, Ferri, DiStasio~Jr, and Tkatchenko]{ambrosetti2016wavelike}
A.~Ambrosetti, N.~Ferri, R.~A. DiStasio~Jr and A.~Tkatchenko, \emph{Science}, 2016, \textbf{351}, 1171--1176\relax
\mciteBstWouldAddEndPuncttrue
\mciteSetBstMidEndSepPunct{\mcitedefaultmidpunct}
{\mcitedefaultendpunct}{\mcitedefaultseppunct}\relax
\EndOfBibitem
\bibitem[Misquitta \emph{et~al.}(2010)Misquitta, Spencer, Stone, and Alavi]{misquitta2010dispersion}
A.~J. Misquitta, J.~Spencer, A.~J. Stone and A.~Alavi, \emph{Phys. Rev. B}, 2010, \textbf{82}, 075312\relax
\mciteBstWouldAddEndPuncttrue
\mciteSetBstMidEndSepPunct{\mcitedefaultmidpunct}
{\mcitedefaultendpunct}{\mcitedefaultseppunct}\relax
\EndOfBibitem
\bibitem[Angyan \emph{et~al.}(2020)Angyan, Dobson, Jansen, and Gould]{AngyanDispersion2020}
J.~Angyan, J.~Dobson, G.~Jansen and T.~Gould, \emph{{London Dispersion Forces in Molecules, Solids and Nano-structures: An Introduction to Physical Models and Computational Methods}}, Royal Society of Chemistry, 2020\relax
\mciteBstWouldAddEndPuncttrue
\mciteSetBstMidEndSepPunct{\mcitedefaultmidpunct}
{\mcitedefaultendpunct}{\mcitedefaultseppunct}\relax
\EndOfBibitem
\bibitem[Mahan(2010)]{MahanBook}
G.~D. Mahan, \emph{{Many-Particle Physics}}, Springer New York, 2010\relax
\mciteBstWouldAddEndPuncttrue
\mciteSetBstMidEndSepPunct{\mcitedefaultmidpunct}
{\mcitedefaultendpunct}{\mcitedefaultseppunct}\relax
\EndOfBibitem
\bibitem[Bostr\"om and Sernelius(1999)]{MBostromCurrDrag_1999}
M.~Bostr\"om and B.~E. Sernelius, \emph{Physica Scripta}, 1999, \textbf{1999}, 89\relax
\mciteBstWouldAddEndPuncttrue
\mciteSetBstMidEndSepPunct{\mcitedefaultmidpunct}
{\mcitedefaultendpunct}{\mcitedefaultseppunct}\relax
\EndOfBibitem
\bibitem[Bostr\"om and Sernelius(2000)]{BostromSerneliusPhysRevB.61.2204}
M.~Bostr\"om and B.~E. Sernelius, \emph{Phys. Rev. B}, 2000, \textbf{61}, 2204--2210\relax
\mciteBstWouldAddEndPuncttrue
\mciteSetBstMidEndSepPunct{\mcitedefaultmidpunct}
{\mcitedefaultendpunct}{\mcitedefaultseppunct}\relax
\EndOfBibitem
\bibitem[Chang \emph{et~al.}(1971)Chang, Cooper, Drummond, and Young]{chang1971van}
D.~Chang, R.~Cooper, J.~Drummond and A.~Young, \emph{Physics Letters A}, 1971, \textbf{37}, 311--312\relax
\mciteBstWouldAddEndPuncttrue
\mciteSetBstMidEndSepPunct{\mcitedefaultmidpunct}
{\mcitedefaultendpunct}{\mcitedefaultseppunct}\relax
\EndOfBibitem
\bibitem[Dobson \emph{et~al.}(2006)Dobson, White, and Rubio]{dobson2006asymptotics}
J.~Dobson, A.~White and A.~Rubio, \emph{{Phys. Rev. Lett.}}, 2006, \textbf{96}, 073201\relax
\mciteBstWouldAddEndPuncttrue
\mciteSetBstMidEndSepPunct{\mcitedefaultmidpunct}
{\mcitedefaultendpunct}{\mcitedefaultseppunct}\relax
\EndOfBibitem
\bibitem[Richmond \emph{et~al.}(1972)Richmond, Davies, and Ninham]{Richmond1972}
P.~Richmond, B.~Davies and B.~Ninham, \emph{Physics Letters A}, 1972, \textbf{39}, 301--302\relax
\mciteBstWouldAddEndPuncttrue
\mciteSetBstMidEndSepPunct{\mcitedefaultmidpunct}
{\mcitedefaultendpunct}{\mcitedefaultseppunct}\relax
\EndOfBibitem
\bibitem[Davies \emph{et~al.}(1973)Davies, Ninham, and Richmond]{Davies1973}
B.~Davies, B.~Ninham and P.~Richmond, \emph{The Journal of Chemical Physics}, 1973, \textbf{58}, 744--750\relax
\mciteBstWouldAddEndPuncttrue
\mciteSetBstMidEndSepPunct{\mcitedefaultmidpunct}
{\mcitedefaultendpunct}{\mcitedefaultseppunct}\relax
\EndOfBibitem
\bibitem[Lechelon \emph{et~al.}(2022)Lechelon, Meriguet, Gori, Ruffenach, Nardecchia, Floriani, Coquillat, Teppe, Mailfert, Marguet, Ferrier, Varani, Sturgis, Torres, and Pettini]{Lechelon2022_sciadv.abl5855}
M.~Lechelon, Y.~Meriguet, M.~Gori, S.~Ruffenach, I.~Nardecchia, E.~Floriani, D.~Coquillat, F.~Teppe, S.~Mailfert, D.~Marguet, P.~Ferrier, L.~Varani, J.~Sturgis, J.~Torres and M.~Pettini, \emph{Science Advances}, 2022, \textbf{8}, eabl5855\relax
\mciteBstWouldAddEndPuncttrue
\mciteSetBstMidEndSepPunct{\mcitedefaultmidpunct}
{\mcitedefaultendpunct}{\mcitedefaultseppunct}\relax
\EndOfBibitem
\bibitem[Ambrosetti \emph{et~al.}(2022)Ambrosetti, Umari, Silvestrelli, Elliott, and Tkatchenko]{AmbrosettiNatCom2022}
A.~Ambrosetti, P.~Umari, P.~L. Silvestrelli, J.~Elliott and A.~Tkatchenko, \emph{Nat. Commun.}, 2022, \textbf{13}, 813\relax
\mciteBstWouldAddEndPuncttrue
\mciteSetBstMidEndSepPunct{\mcitedefaultmidpunct}
{\mcitedefaultendpunct}{\mcitedefaultseppunct}\relax
\EndOfBibitem
\bibitem[Parsegian(2006)]{paresegian2006}
V.~A. Parsegian, \emph{Van der Waals forces}, Cambridge University Press, 2006\relax
\mciteBstWouldAddEndPuncttrue
\mciteSetBstMidEndSepPunct{\mcitedefaultmidpunct}
{\mcitedefaultendpunct}{\mcitedefaultseppunct}\relax
\EndOfBibitem
\bibitem[Richmond and Davies(1972)]{richmond1972many}
P.~Richmond and B.~Davies, \emph{Molecular Physics}, 1972, \textbf{24}, 1165--1168\relax
\mciteBstWouldAddEndPuncttrue
\mciteSetBstMidEndSepPunct{\mcitedefaultmidpunct}
{\mcitedefaultendpunct}{\mcitedefaultseppunct}\relax
\EndOfBibitem
\bibitem[Smith \emph{et~al.}(1973)Smith, Mitchell, and Ninham]{smith1973van}
E.~Smith, D.~Mitchell and B.~Ninham, \emph{Journal of Theoretical Biology}, 1973, \textbf{41}, 149--160\relax
\mciteBstWouldAddEndPuncttrue
\mciteSetBstMidEndSepPunct{\mcitedefaultmidpunct}
{\mcitedefaultendpunct}{\mcitedefaultseppunct}\relax
\EndOfBibitem
\bibitem[Martinov and Nikolov(1971)]{martinov1971structure}
N.~K. Martinov and N.~A. Nikolov, \emph{Journal of Physics A}, 1971, \textbf{4}, 464\relax
\mciteBstWouldAddEndPuncttrue
\mciteSetBstMidEndSepPunct{\mcitedefaultmidpunct}
{\mcitedefaultendpunct}{\mcitedefaultseppunct}\relax
\EndOfBibitem
\bibitem[Davies and Ninham(1972)]{davies1972van}
B.~Davies and B.~W. Ninham, \emph{The Journal of Chemical Physics}, 1972, \textbf{56}, 5797--5801\relax
\mciteBstWouldAddEndPuncttrue
\mciteSetBstMidEndSepPunct{\mcitedefaultmidpunct}
{\mcitedefaultendpunct}{\mcitedefaultseppunct}\relax
\EndOfBibitem
\bibitem[Stratton(1941)]{stratton1941}
J.~A. Stratton, \emph{Electromagnetic Theory}, Mcgraw Hill Book Company, 1941\relax
\mciteBstWouldAddEndPuncttrue
\mciteSetBstMidEndSepPunct{\mcitedefaultmidpunct}
{\mcitedefaultendpunct}{\mcitedefaultseppunct}\relax
\EndOfBibitem
\bibitem[Watson(1958)]{watson1958}
G.~N. Watson, \emph{Treatise on the Theory of Bessel Functions}, Cambridge University Press, 1958\relax
\mciteBstWouldAddEndPuncttrue
\mciteSetBstMidEndSepPunct{\mcitedefaultmidpunct}
{\mcitedefaultendpunct}{\mcitedefaultseppunct}\relax
\EndOfBibitem
\bibitem[Ninham \emph{et~al.}(1970)Ninham, Parsegian, and Weiss]{NinhamParsegianWeiss1970}
B.~W. Ninham, V.~A. Parsegian and G.~H. Weiss, \emph{J. Stat. Phys.}, 1970, \textbf{2}, 323\relax
\mciteBstWouldAddEndPuncttrue
\mciteSetBstMidEndSepPunct{\mcitedefaultmidpunct}
{\mcitedefaultendpunct}{\mcitedefaultseppunct}\relax
\EndOfBibitem
\bibitem[Bostr\"om \emph{et~al.}(2024)Bostr\"om, Gholamhosseinian, Pal, Li, and Brevik]{physics6010030}
M.~Bostr\"om, A.~Gholamhosseinian, S.~Pal, Y.~Li and I.~Brevik, \emph{Physics}, 2024, \textbf{6}, 456--467\relax
\mciteBstWouldAddEndPuncttrue
\mciteSetBstMidEndSepPunct{\mcitedefaultmidpunct}
{\mcitedefaultendpunct}{\mcitedefaultseppunct}\relax
\EndOfBibitem
\bibitem[Mahanty and Ninham(1976)]{Maha}
J.~Mahanty and B.~W. Ninham, \emph{{Dispersion Forces}}, Academic Press, London, 1976\relax
\mciteBstWouldAddEndPuncttrue
\mciteSetBstMidEndSepPunct{\mcitedefaultmidpunct}
{\mcitedefaultendpunct}{\mcitedefaultseppunct}\relax
\EndOfBibitem
\bibitem[Smith \emph{et~al.}(1973)Smith, Mitchell, and Ninham]{SmithMitchellNinham1970}
E.~Smith, D.~Mitchell and B.~Ninham, \emph{Journal of Theoretical Biology}, 1973, \textbf{41}, 149--160\relax
\mciteBstWouldAddEndPuncttrue
\mciteSetBstMidEndSepPunct{\mcitedefaultmidpunct}
{\mcitedefaultendpunct}{\mcitedefaultseppunct}\relax
\EndOfBibitem
\bibitem[Dzyaloshinskii \emph{et~al.}(1961)Dzyaloshinskii, Lifshitz, and Pitaevskii]{Dzya}
I.~Dzyaloshinskii, E.~Lifshitz and L.~Pitaevskii, \emph{Adv. Phys.}, 1961, \textbf{10}, 165--209\relax
\mciteBstWouldAddEndPuncttrue
\mciteSetBstMidEndSepPunct{\mcitedefaultmidpunct}
{\mcitedefaultendpunct}{\mcitedefaultseppunct}\relax
\EndOfBibitem
\bibitem[Craig and Thirunamachandran(1982)]{Craig}
D.~P. Craig and T.~Thirunamachandran, \emph{Adv. Quant. Chem.}, 1982, \textbf{16}, 97\relax
\mciteBstWouldAddEndPuncttrue
\mciteSetBstMidEndSepPunct{\mcitedefaultmidpunct}
{\mcitedefaultendpunct}{\mcitedefaultseppunct}\relax
\EndOfBibitem
\bibitem[Parsegian(2006)]{Pars}
V.~A. Parsegian, \emph{Van der Waals forces: A handbook for biologists, chemists, engineers, and physicists}, Cambridge University Press, New York, 2006\relax
\mciteBstWouldAddEndPuncttrue
\mciteSetBstMidEndSepPunct{\mcitedefaultmidpunct}
{\mcitedefaultendpunct}{\mcitedefaultseppunct}\relax
\EndOfBibitem
\bibitem[Ninham \emph{et~al.}(1969)Ninham, Nossal, and Zwanzig]{NinhamNossalZwanzig1969}
B.~W. Ninham, R.~Nossal and R.~Zwanzig, \emph{J. Chem. Phys.}, 1969, \textbf{51}, 5028--5033\relax
\mciteBstWouldAddEndPuncttrue
\mciteSetBstMidEndSepPunct{\mcitedefaultmidpunct}
{\mcitedefaultendpunct}{\mcitedefaultseppunct}\relax
\EndOfBibitem
\bibitem[Bostr\"om \emph{et~al.}(2002)Bostr\"om, Williams, and Ninham]{MathiasDavidBarryDNA2002}
M.~Bostr\"om, D.~R.~M. Williams and B.~W. Ninham, \emph{J. Phys. Chem.}, 2002, \textbf{106}, 7908--7912\relax
\mciteBstWouldAddEndPuncttrue
\mciteSetBstMidEndSepPunct{\mcitedefaultmidpunct}
{\mcitedefaultendpunct}{\mcitedefaultseppunct}\relax
\EndOfBibitem
\bibitem[Hyde \emph{et~al.}(1997)Hyde, Ninham, Andersson, Larsson, Landh, Blum, and Lidin]{HydeLanuage}
S.~Hyde, B.~W. Ninham, S.~Andersson, K.~Larsson, T.~Landh, Z.~Blum and S.~Lidin, \emph{{The Language of Shape}}, Elsevier Science B.V., Amsterdam, 1997, pp. 237--256\relax
\mciteBstWouldAddEndPuncttrue
\mciteSetBstMidEndSepPunct{\mcitedefaultmidpunct}
{\mcitedefaultendpunct}{\mcitedefaultseppunct}\relax
\EndOfBibitem
\end{mcitethebibliography}

\end{document}